\documentclass[twocolumn,english,prd,superscriptaddress,nofootinbib,preprintnumbers]{revtex4}

\makeatletter

\usepackage{amssymb}
\usepackage{amsmath}
\usepackage{dcolumn}
\usepackage[latin1]{inputenc}

\usepackage{ifpdf}
\ifpdf 
  \usepackage[pdftex]{graphicx}
  \usepackage{pgf}
  \usepackage{tikz}
\else 
  \usepackage{graphicx}
  \DeclareGraphicsExtensions{.eps,.png}
  \usepackage{pgf}
  \usepackage{tikz}
  \usepackage{pstricks}
\fi

\makeatother

\begin{document}

\title{Dark Interactions and Cosmological Fine-Tuning}

\author{Miguel Quartin}
\email{mquartin@if.ufrj.br}

\author{Maurício O. Calvão}
\email{orca@if.ufrj.br}

\author{Sergio E. Jorás}
\email{joras@if.ufrj.br}

\author{Ribamar R. R. Reis}
\email{ribamar@if.ufrj.br}

\author{Ioav Waga}
\email{ioav@if.ufrj.br}

\affiliation{Universidade Federal do Rio de Janeiro, Instituto de Física, CEP 21941-972, Rio
de Janeiro, RJ, Brazil}

\date{\today{}}

\begin{abstract}
    Cosmological models involving an interaction between dark matter and dark energy have been proposed in order to solve the so-called coincidence problem. Different forms of coupling have been studied, but there have been claims that observational data seem to narrow (some of) them down to something annoyingly close to the $\Lambda$CDM model, thus greatly reducing their ability to deal with the problem in the first place. The smallness problem of the initial energy density of dark energy has also been a target of cosmological models in recent years. Making use of a moderately general coupling scheme, this paper aims to unite these different approaches and shed some light as to whether this class of models has any true perspective in suppressing the aforementioned issues that plague our current understanding of the universe, in a quantitative and unambiguous way.
\end{abstract}
\maketitle

\section{Introduction}\label{sec:intro}

Several sets of observational data such as Type Ia supernovae (SNIa), Cosmic Microwave Background (CMB) and large scale structure surveys, when combined, indicate that we live in a nearly flat, low-matter-density ($\Omega_{m0} \sim 0.3$) universe whose expansion is speeding up at present. The driving source of this cosmic acceleration is nevertheless still poorly understood. The simplest candidate is Einstein's cosmological constant $\Lambda$, assumed to stand for the vacuum energy density ($\rho_\Lambda = \Lambda/8\pi G$). However its tiny value, as inferred from observations, is some $50$ ($120$) orders of magnitude below conservative (aggressive) estimates given by quantum field theory and explaining this discrepancy is one of the most difficult problems in theoretical physics. This constitutes what is sometimes called the \emph{cosmological constant problem} (CCP), whose solution probably require the discovery of a yet unknown underlying symmetry capable of inducing a nearly perfect cancellation of the vacuum energy density value (candidates such as supersymmetry do not help since they are broken at high energy scales) or at least of its contribution to gravity. Such fine-tuned cancellation is thought to be very unlikely, and the problem is usually addressed by what is considered a more reasonable hypothesis: a complete suppression of the vacuum energy density contribution to the gravitational sector. This in turn requires a different explanation for the aforementioned cosmological observations.

Possible candidates discussed in the literature usually fall into two categories: exotic components with negative pressure (dark energy or simply DE) or proper modifications of general relativity which become relevant at cosmological scales. In either case different observational and theoretical problems arise, and this made the soil fertile for a plethora of models to bloom. On the theoretical side we identify three possible issues which we will dub: the \emph{coincidence problem} (CP), the \emph{DE energy density initial condition problem} (DEICP) and the \emph{sensitivity to initial conditions} (SIC). These can be strongly related (as in the case of any $\Lambda$-like dark energy model, defined below) or even dependent on the physical interpretation underlying a particular model (as opposed to being a problem of the model itself), as we will clarify below. Another important point is that all three problems often impose \emph{fine-tuning} to observationally viable models.

The CP can be stated as: why only recently (in terms of redshift) has the DE energy density become comparable to the matter one, since both components are usually assumed independent and thus scale in different ways? The \linebreak[4] DEICP appears when one interprets dark energy as a proper field, but not when one works in modified gravity theories. In other words: why is the initial (by which we mean just after inflation, $z \sim 10^{26}$) value of the DE energy density much smaller than what we would expect from equipartition considerations? It seems natural to expect that after inflation the different fields in nature would have energy densities with the same order of magnitude. Thus a reasonable value for the ratio of energy densities between DE and radiation,  $\rho_{\rm DE} / \rho_{\rm r}$ , would be $\thicksim 10^{-2}$ or $10^{-3}$. The SIC is a measure of the robustness of a model to different initial conditions, which usually can be thought in terms of basins of attraction. Models with a larger SIC require a more fine-tuned initial condition than those with a smaller one. The amount of SIC is not always considered an issue, as the initial condition for radiation itself can be thought of as fine-tuned in order to match present measurements of the CMB temperature. Nevertheless, models which allow a broader range of initial conditions are usually preferred over those which put stringent constraints on such a range. The relevance of each of these three issues (CP, DEICP and SIC) can be disputed, but our main intention here is to clarify their distinction, as is not uncommon for ambiguity among them (and even with the CCP) to arise in the literature.

Before we proceed, a remark on notation is in order. To make explicit our phenomenological approach (which includes for instance modified gravity models) in our discussions we will denote by the indexes ``$_x$'' and ``$_{\rm DE}$'' all physical quantities related to DE, but we will reserve the latter exclusively to the cases in which only the field interpretation makes sense. Furthermore, we will use the index suffixes ``$_0$'' and ``$_i$'' to denote respectively quantities evaluated today and at the end of inflation. On an altogether different issue, we will refer to $1\sigma$, $2\sigma$ and $3\sigma$ as a shorthand notation for $68.3\%$, $95.4\%$ and $99.73\%$ confidence levels, respectively, even though this would only be rigorously the case for gaussian likelihoods.


In $\Lambda$CDM (henceforth $\Lambda$CDM will denote any model which behaves like a cosmological constant but which is not described by the vacuum energy density) or in any other constant-$w_x$ CDM model, the CP and DEICP cannot be jointly addressed, and either one requires a fine-tuning of a parameter (e.g. $\rho_{\rm DEi}$). To have a $\Lambda$ dominated universe today, the initial DE energy density has to be set dozens of orders of magnitudes smaller than the initial matter energy density. However this may not be the case, for instance, for quintessence tracker scalar field models~\cite{Steinhardt98}. Attractor-like solutions appear in this case such that, for a wide range of initial conditions, the field eventually reaches a definite cosmic evolutionary track at which $\Omega_{m0} \sim \Omega_{\rm DE0}$. Assuming that the background energy density decreases with the scale factor as $\rho_{\rm back} \propto a^{-n}$, we call tracker those solutions for which the energy density of the tracker field scales as $\rho_{x} \propto a^{-m}$, where $m$ depends on the equation of state of the background but is such that $m < n$. \emph{Scaling}, on the other hand, occurs when the ratio $\rho_{x}/\rho_{m}$ is constant \cite{Ferreira97}, that is, when $m=n$. Adjusting the onset of tracking is one of the difficulties of quintessence tracker solutions. For instance, starting from equipartition conditions after inflation it was shown in~\cite{Steinhardt98} that, in inverse power law tracking quintessence models ($V(\phi)\propto \phi^{-\alpha}$), only for $\alpha>5$ will the field start tracking before matter-radiation equality. However current observations impose $\alpha \lesssim 1$~\cite{Wilson06,Waga00}, and in this case the field would only start tracking at recent epochs. Since DE must develop negative pressure around $z \thicksim 1$ in order to drive acceleration, this demands a fine-tuning to the field potential. Therefore, although the DEICP can be addressed by these models, they pose no clear advantage as far as the CP is concerned, as the latter could be reformulated as: why do we live in a special epoch when the field just started tracking? In order to circumvent this difficulty, models with a non-canonical kinetic term (k-essence)~\cite{Chiba00,Picon00} have been proposed as an alternative. In k-essence, the field reaches a scaling solution during the radiation dominated era, but dynamics (rather than an adjustment of a parameter) triggers a transition when non-relativistic matter starts to dominate, after which k-essence follows an accelerated attractor. However, a good candidate with a large basin of attraction is still lacking~\cite{Malquarti03}.

Here we shall follow the approach of~\cite{AQTW}, and address the CP by requiring a duo-scaling cosmology. We seek models whose phase space exhibit (at least) two fixed points: one responsible for the matter dominated era (henceforth MDE) and the other for the present dark energy dominated acceleration. The former needs to be a saddle point and the latter (preferably) an attractor. The CP could be solved by a model in which our present universe has reached (by which we mean it is close enough for a given criterium) such an accelerated attractor, as this would mean that the current cosmological energy distribution is not a transient phase but rather an unavoidable and permanent regime. The existence of such stable and accelerated fixed point requires a coupling between matter and dark energy \cite{Amendola00}, as otherwise the DE energy density would have to decrease with $a^{-3}$.

In order to quantify both the DEICP and SIC, we will define two quantities: $\,\zeta\equiv \rho_{{\rm DE}i}^{\rm max} / \rho_{ri}\,$ and \linebreak[4] $\,\Delta\equiv (\rho_{xi}^{\rm max} - \rho_{xi}^{\rm min}) / \left|\rho_{xi}^{\rm max}\right|  \,$, where $\rho_r$ stands for the radiation energy density and $\rho_{xi/{\rm DE}i}^{\rm max/min}$ denotes the maximum and minimum values of $\rho_{xi/{\rm DE}i}$ that evolve to the present day observed values within~$1\sigma$. Models which have small SIC are characterized by large values of~$\Delta$ and vice-versa. Note that our choice of variable for measuring the SIC is a ratio of the size of the permissible region to the maximum value of that range, and thus is not directly related with the proper size of the allowed region of initial conditions. In other words, $\Delta$ is a relative measure, not an absolute one. One is  sometimes more interested in evaluating the absolute width of the initial condition range and in this case $(\rho_{xi}^{\rm max} - \rho_{xi}^{\rm min})$ could be used instead. One should bear in mind that newer and improved observations will probably narrow our present uncertainties on $\rho_x$ and this could reflect on the initial range $\Delta$, which should depend only on the model and not on the quality of our observations. This is avoided by computing the ratio $\Delta / \Delta_{\Lambda \textrm{CDM}}$. Furthermore, our definition of $\Delta$ is not a good one in the cases where $\rho_{xi}^{\rm max}$ is very close to zero and $\rho_{xi}^{\rm min}$ is not.  Finally, for the CP we will use the ratio $R_z$ between DE and dark matter energy densities, where $z$ stands for the redshift, as a measure of how close we are to the accelerated attractor. More specifically, a solution to the CP requires $R_{-1} \simeq R_0$.

Making use of a three-parameter model with a coupling scheme that generalizes some previous ones in the literature (section~\ref{sec:DIM}), we analyse the phase space (section~\ref{sec:fpoints}) and apply two different cosmological tests (section~\ref{sec:obs-constraints}): SNIa, as given by a combined catalog~\cite{Davis07}, and the so-called CMB shift parameter, as inferred from WMAP3~\cite{Wang07}. We look at three aspects: what is the parametric region that allows such duo-scaling cosmology (section~\ref{sec:2epochs}); how does the model cope with the DEICP inside that parametric region; and what is the SIC (section~\ref{sec:fine-tuning}). We also propose a two-parameter toy model (section~\ref{sec:toy-model}) for which there exists analytic solutions and which has small SIC, alleviate the CP and, in some cases, also the DEICP. Finally, we note that by choosing freely all three parameters we can actually solve the CP.

\section{The Dark Interactions Model}\label{sec:DIM}

We consider the universe as filled by four components: radiation, baryons, and two coupled barotropic (latu sensu) fluids. One, a cold, pressureless, dark matter (CDM) and the other a negative pressure dark energy. We will denote them, respectively, by the subindexes $r$, $b$, $c$ and $x$. For the sake of simplicity, our analysis will be restricted to models with constant equation of state parameter $w_x \equiv p_x/\rho_x$. Local gravity constraints limit greatly any possible interactions between dark energy and baryons. To allow such a coupling requires a mechanism to either make these constraints only effective in the present and not in the past, or to make the interaction range short enough, as achieved in the chameleon scalar field models~\cite{Mota04,Khoury04}. We therefore neglect this possibility and focus on a coupling with dark matter alone. An interaction with radiation is also discarded on the grounds that any such reaction would not affect the dynamics of the system near the sought two scaling regimes, required to address the CP. On the other hand, a coupling with radiation might be desirable if one wants to address also the DEICP. Since this would introduce many difficulties of its own, we will not consider it further in this work.

Our interest is in cosmological scaling solutions in a spatially flat Friedmann-Lemaître-Robertson-Walker (FLRW) background metric with a scale factor $a(t)$:
\begin{equation}
    {\textrm{d}}s^{2}=-{\textrm{d}}t^{2}+a^{2}(t){\textrm{d}}\mathbf{x}^{2}. \label{eq:FRW}
\end{equation}
The Friedmann equation in General Relativity is given by
\begin{equation}
    3H^{2}=M_{P}^{-2}(\rho_x+\rho_c+\rho_b+\rho_r)\,,\label{eq:Friedmann}
\end{equation}
where $M_{P}^{-2}=8\pi G$ with $G$ being the gravitational constant, and where the different $\rho_i$ make up the energy density of the universe. In what follows we shall set $M_{P}=1$. Different forms of coupling have been considered in the literature. We may take the coupling between CDM and DE to be such that
\begin{equation}\label{eq:cons-eq-field-c-x}
\begin{aligned}
    \frac{{\textrm{d}}\rho_x}{{\textrm{d}}N}+3(1+w_x)\rho_{x}&= -3 Q(\rho_c,\rho_x),  \\
    \frac{{\textrm{d}}\rho_c}{{\textrm{d}}N}+3\rho_{c}&= 3 Q(\rho_c,\rho_x),
\end{aligned}
\end{equation}
where $N\equiv \ln a$, sometimes called the number of $e$-folds. Then, some of the (non-mutually exclusive) proposed forms for $Q(\rho_c,\rho_x)$ so far are:
\begin{equation}
    Q(\rho_c,\rho_x) =
    \left\{
    \begin{aligned}
        &(\overline{\lambda} - w_x) \, \frac{\rho_x \rho_c}{\rho_x + \rho_c}& \text{\cite{Cai05,Zimdahl03}}, \\
        &\overline{\lambda} \, \frac{\rho_c}{\sqrt{\rho_x + \rho_c}}& \text{\cite{Boehmer08}}, \\
        &\overline{\lambda} \, (\rho_x + \rho_c) & \text{\cite{Chimento03}}, \\
        &\overline{\lambda} \, \rho_c & \text{\cite{Zimdahl01,Guo07}}, \\
        &\overline{\lambda}_1 \, \rho_x^{\overline{\lambda}_2} \rho_c^{\overline{\lambda}_3} & \text{\cite{Mangano03}},
    \end{aligned}
    \right.
\end{equation}
where, in addition to $w_x$, we take $\overline{\lambda}$ and the different $\overline{\lambda}_i$ to be constants. Note that positive values of $Q(\rho_c,\rho_x)$ indicate that energy is being transferred from dark energy to dark matter, meaning that the latter will dilute slower than in the case without interactions.

Our analysis will be restricted to the case\footnote{It has only recently come to our attention that this proposed form was previously considered by \cite{Sadjadi06}.}
\begin{eqnarray}\label{eq:coupling-form}
    Q(\rho_c,\rho_x) = \lambda_x \rho_x + \lambda_c \rho_c,
\end{eqnarray}
which is a more general form than those found in~\cite{Chimento03} or~\cite{Guo07}, but reduces to those cases, respectively, whenever $\lambda_x = \lambda_c \equiv \overline{\lambda}\;$ and $\;\lambda_x = 0$, $\lambda_c\equiv \overline{\lambda}$. It is, however, in a different class than the ones used in many interacting scalar field models. In particular, it is not equivalent to the coupling used in~\cite{AQTW} and therefore it might circumvent the tough challenges regarding the solution of the CP imposed there. Furthermore, in what follows we will always consider $w_x$ to be a negative constant, whose value is the third free parameter of our model.

One of the reasons for considering the above coupling scheme is that it is a straightforward extension of some of the other forms encountered in the literature. Nonetheless, it has some distinctive features that the reader should bear in mind. First, when one introduces a coupling term which is proportional to $\rho_c$ but not, say, $\rho_c \rho_x$, one has to be careful about what one means by ``dark matter''. The reason is that even in the absence of dark energy, the former has a non zero equation of state parameter. In fact, for $\,\rho_x=0\,$ we have $\,w_c^{\rm eff} = -\lambda_c$, as can be easily seen from eq.~\ref{eq:cons-eq-field-c-x}. Second, such coupling naturally allows for $\rho_x$ and $\rho_c$ to become negative as the universe evolves, and the way to deal with this  is discussed in the following section. Third, the overall sign of $Q(\rho_c,\rho_x)$ may change as the energy densities dilute. A late-time scaling leading to cosmological acceleration enforces a positive value today, but leaves open the possibility of an opposite energy transfer in the past, which in turn could lead to a stronger formation of structures during the MDE.

Rewriting eqs.~\eqref{eq:cons-eq-field-c-x} using eq.~\eqref{eq:coupling-form} we get
\begin{equation} \label{eq:cons-eq-field-c-x-our-Q}
\begin{aligned}
    &\frac{{\textrm{d}}\rho_x}{{\textrm{d}}N}+3(1+w_x+\lambda_x)\rho_{x}=
    -3\,\lambda_c\,\rho_{c},\\
    &\frac{{\textrm{d}}\rho_c}{{\textrm{d}}N}+3(1-\lambda_c)\rho_{c}= 3\,\lambda_x\,\rho_{x}.
\end{aligned}
\end{equation}
This system does not admit analytic solutions in the general case\footnote{General solutions exist if one neglects baryons and radiation, as has been shown in~\cite{Barrow06}.} but there are noteworthy exceptions for some particular ones, namely: (i)~$\lambda_x = 0$; (ii)~$\lambda_c = 0$; (iii)~$\lambda_c = 1\,$ and $\,$(iv)~$w_x = -1 - \lambda_x$. For each of these possibilities, it is possible to decouple the above equations. Case~(i) is the one studied in~\cite{Guo07}. The second case, while in principle worth investigating, turns up to be incapable of alleviating the initial condition problem, as we will show in section~\ref{sec:fine-tuning}. Case~(iii) is completely ruled out by observations, as will become clear in section~\ref{sec:obs-constraints}. The last possibility, however, is an interesting one. It allows simultaneously $w_x$ and $\lambda_x$ to be close to $\,-1\,$ and $\,0$, respectively, while still exhibiting a distinct behaviour compared to $\Lambda$CDM. Not only can it easily satisfy the observational tests we employed, but it also addresses both the CP and DEICP. We will return to this particular choice of parameters in section~\ref{sec:toy-model}.

\section{Fixed points}\label{sec:fpoints}

In order to study the dynamics of the model, we shall introduce the following set of variables,
\begin{equation}
\begin{aligned}
    X&=\frac{1}{H^2} \frac{\rho _x}{3},   &\;\;Y&=\frac{1}{H^2}\frac{\rho _c}{3}, \\
    z&=\frac{1}{H}\sqrt{\frac{\rho _b}{3}},   &\;\;u&=\frac{1}{H}\sqrt{\frac{\rho _r}{3}},
\end{aligned}
\end{equation}
and rewrite the continuity equations and eq.~(\ref{eq:Friedmann}) as
\begin{equation} \label{eq:autonomous}
\begin{aligned}
    &\frac{{\textrm{d}}X}{{\textrm{d}}N} \;=\; X \left[3 w_x (X-1) + u^2 - 3\lambda_x \right] - 3\lambda_c \,Y, \\
    &\frac{{\textrm{d}}Y}{{\textrm{d}}N} \;=\; Y \left[3 w_x X + u^2 + 3\lambda_c \right] + 3\lambda_x X, \\
    &\frac{{\textrm{d}}z}{{\textrm{d}}N} \;=\; \frac{z}{2} \left[3 w_x X + u^2     \right], \\
    &\frac{{\textrm{d}}u}{{\textrm{d}}N} \;=\; \frac{u}{2} \left[3 w_x X + u^2 - 1 \right].
\end{aligned}
\end{equation}
Such a set of equations, together with the constraint
\begin{equation} \label{eq:constraint-4d}
    X+Y+z^2+u^2=1
\end{equation}
imposed by eq.~(\ref{eq:Friedmann}), form the autonomous system upon which the remainder of this paper will be based. Two remarks are in order here. First, we define $X$ and $Y$ and not, say, $x^2$ and $y^2$ as variables to naturally allow for negative $\rho_x$ and $\rho_c$. While $\,\rho_x < 0\,$ not only cannot be ruled out but in fact is mandatory in many $f(R)$ theories that aim to solve the coincidence problem and satisfy current observational constraints~\cite{Amendola07-PhCross}, one could argue that $\,\rho_c < 0\,$ is unphysical or at least undesirable. Nevertheless this class of models demands such flexibility, and to deny it a priori leads to an incomplete understanding of the dynamics involved.\footnote{A previous work \cite{Olivares07} neglected this, and in so doing overlooked the existence of some fixed points and allowed for trajectories which would eventually violate the condition (\ref{eq:constraint-4d}).} Any such physical considerations must be done a posteriori, for instance, by ruling out trajectories for which $\,\rho_c < 0\,$ at some point. Second, due to the above constraint the system reduces to a $3-$dimensional one, and therefore one of the equations~(\ref{eq:autonomous}) may be eliminated. By doing so the reduced system still has to satisfy a constraint which, choosing to eliminate $z$, is written as
\begin{equation} \label{eq:constraint-3d}
    X+Y+u^2\leq 1.
\end{equation}

Our notation for the fixed points will be in the form $\,(X_{\rm fp},Y_{\rm fp},z_{\rm fp},u_{\rm fp})$. From the last two equations of system~(\ref{eq:autonomous}) it is straightforward to see that, for any fixed point, $\,z_{\rm fp}\neq 0\,$ implies $\,u_{\rm fp}=X_{\rm fp}=0\,$. This in turn forces either $\,Y_{\rm fp}=0\,$ or $\,\lambda_c=0\,$. The former option corresponds to the baryonic point $\,(0,0,1,0)\,$. The latter, to a regular CDM universe which is known to converge to $\,(0,Y_{\rm fp},z_{\rm fp},0)\,$. All other fixed points will have $\,z_{\rm fp}=0\,$. On the other hand, when $\,u_{\rm fp}\neq 0\,$ it can be shown that the only possible fixed point is $\,(0,0,0,1)\,$, unless $\lambda_x ,\lambda_c$ and $w_x$ are chosen to match a very specific relation given by
\begin{equation}\label{eq:condition-a-b-w}
    (1-3w_x) (1 + 3\lambda_c ) = 3\lambda_x.
\end{equation}
Thus, the fixed points of interest, responsible for the MDE and present acceleration, can be written as $\,(X_{\rm fp},Y_{\rm fp},0,0)\,$. As mentioned before, we shall follow the approach of~\cite{AQTW} and seek two such points: a saddle point responsible for the MDE, and an attractor governing the dark energy dominated epoch.

\begin{table*}[t]
\begin{center}\begin{tabular}{|c|c|c|c|c|c|c|c|}
    \hline \textbf{Point} & $\mathbf{X}$ & $\mathbf{Y}$& $\,\mathbf{z}\,$& $\,\mathbf{u}\,$ & \textbf{Existence}
    & \textbf{Stability} & \textbf{Acceler.} \tabularnewline
    \hline
    $A_1$ & $X_{A} < 0$     & $1-X_{A}$ &0&0& \parbox[c][1.4cm]{7.0cm}{$\,\lambda_c < 0 $ \linebreak \fbox{\textbf{or}} \linebreak $\,\lambda_c > 0 \;$ and $\,\lambda_x > \lambda_c \;\,\mbox{ and }\; |w_x| < \left(\sqrt{\lambda_x} - \sqrt{\lambda_c}\right)^2\,$} & \parbox[c][1cm]{3cm}{saddle under condition (\ref{eq:cond-saddle-XA1})
     \linebreak unstable otherwise} & no \tabularnewline
     \hline
    $A_2$ & $0\leq X_{A}<1$ & $1-X_{A}$ &0&0& \parbox[c][1.4cm]{7cm}{$\,\lambda_x < 0 \;$ and $\,\lambda_c > 0$
     \linebreak \fbox{\textbf{or}} \linebreak $\,\lambda_x,\lambda_c \geq 0 \;\,\mbox{ and }\;
     |w_x| > \left(\sqrt{\lambda_x} + \sqrt{\lambda_c}\right)^2\,$} & saddle point & $w_x X_A < -\frac{1}{3}$ \tabularnewline
     \hline
    $A_3$ & $X_{A} > 1$     & $1-X_{A}$ &0&0& \parbox[c][.7cm]{7cm}{$\,0< \lambda_x < \lambda_c  \,\mbox{ and }\, |w_x| < \left(\sqrt{\lambda_x} - \sqrt{\lambda_c}\right)^2\,$}  & saddle point & $w_x X_A < -\frac{1}{3}$ \tabularnewline
     \hline
    $B_1$ & $X_{B} < 0$     & $1-X_{B}$ &0&0& \parbox[c][.7cm]{7cm}{$\,\lambda_c > 0 \;$ and $\,\lambda_x > \lambda_c \; \mbox{ and }\, |w_x| < \left(\sqrt{\lambda_x} - \sqrt{\lambda_c}\right)^2\,$}  & saddle point & no \tabularnewline
     \hline
    $B_2$ & $0<X_{B}\leq 1$ & $1-X_{B}$ &0&0& \parbox[c][1.4cm]{7cm}{$\,\lambda_x > 0 \;\,$ and $\;\lambda_c < 0$
     \linebreak \fbox{\textbf{or}} \linebreak $\,\lambda_x,\lambda_c \geq 0 \;\,\mbox{ and }\;  |w_x| > \left(\sqrt{\lambda_x} + \sqrt{\lambda_c}\right)^2\,$} & attractor & $w_x X_B < -\frac{1}{3}$ \tabularnewline
     \hline
    $B_3$ & $X_{B} > 1$     & $1-X_{B}$ &0&0& \parbox[c][1.4cm]{7cm}{$\,\lambda_x < 0 \;$ \linebreak \fbox{\textbf{or}}
     \linebreak $\,0< \lambda_x < \lambda_c \,\mbox{ and }\, |w_x| < \left(\sqrt{\lambda_x} - \sqrt{\lambda_c}\right)^2$} & attractor & $w_x X_B < -\frac{1}{3}$ \tabularnewline
     \hline
    $C$ &  \parbox[c][.7cm]{.6cm}{0}& 0&$\;1\;$&$\;0\;$& $\forall\, {w_x, \lambda_x, \lambda_c}$ & saddle point & no \tabularnewline
     \hline
    $D$ &  \parbox[c][.7cm]{.6cm}{0}& 0&0&1& $\forall\, {w_x, \lambda_x, \lambda_c}$ & unstable or saddle & no \tabularnewline
     \hline
    $E$ &  \parbox[c][.7cm]{.6cm}{0}& $Y_E$ & $z_E$ &0& $ \lambda_c = 0 $ & \parbox[c][1cm]{3cm}{saddle for $|w_x|>\lambda_x$
     \linebreak attractor otherwise} & no \tabularnewline
     \hline
    $F$ &  \parbox[c][.7cm]{.6cm}{$-\infty$}& $+\infty$ &$+\infty$&$+\infty$& undetermined, but see figure \ref{fig:Contours-wx=-1} & N/A  & N/A \tabularnewline
     \hline
\end{tabular}\end{center}

\caption{The properties of critical points for the Dark Interactions Model. It is implicitly assumed that $w_x<0$ and that condition~(\ref{eq:condition-a-b-w}) is \emph{not} met. The last column show the conditions for late-time acceleration of the universe. Figure~\ref{fig:phasespace} illustrates this table.}

\label{tab:fixed-pts}
\end{table*}

These two important points will be dubbed \linebreak[4] $\,A\equiv(X_A,Y_A,0,0)\,$ and $\,B\equiv(X_B,Y_B,0,0)\,$. From the system
(\ref{eq:autonomous}) we find
\begin{equation}
    X_A = \frac{w_x + \lambda_x - \lambda_c + \sqrt{(w_x + \lambda_x - \lambda_c )^2 + 4 w_x \lambda_c }}{2 w_x},
\end{equation}
\begin{equation}
    X_B = \frac{w_x + \lambda_x - \lambda_c - \sqrt{(w_x + \lambda_x - \lambda_c )^2 + 4 w_x \lambda_c
    }}{2 w_x},
\end{equation}
with $\,Y_{A,B} = 1-X_{A,B}\,$. Note that $X_B \geq X_A$, since $w_x<~\!0$. In section~\ref{sec:2epochs} we will show that points $A$ and $B$ are candidates for the MDE and present accelerated epoch, respectively. For the moment, it is useful to introduce the inequalities
\begin{align}
    -w_x = |w_x| > \left(\sqrt{\lambda_x} + \sqrt{\lambda_c}\right)^2\,, \label{eq:condition1-on-w}\\
    -w_x = |w_x| < \left(\sqrt{\lambda_x} - \sqrt{\lambda_c}\right)^2 \,. \label{eq:condition2-on-w}
\end{align}
Note that the second condition on $w_x$ is very strong since an eventual acceleration requires either $\lambda_x$ or $\lambda_c$ to be order unity for $w_x$ to be negative enough, which is at odds with what we expect from previous results~\cite{Olivares07}. Both points exist whenever: (i) $\,\lambda_x >0$, $\,\lambda_c >0\,$ and either condition~(\ref{eq:condition1-on-w}) or~(\ref{eq:condition2-on-w}) is met; or (ii) at least one of $\lambda_x$ and $\lambda_c$ is negative. However, depending on the sign of these parameters we can have $X_{A,B}$ negative or larger than~1. If $\,\lambda_x \mbox{ and } \lambda_c\,$ are positive and condition
(\ref{eq:condition1-on-w}) is met, both points are placed between 0 and 1. On the other hand if the inequality (\ref{eq:condition2-on-w}) is the one satisfied, both $X_A$ and $X_B$ will be negative (for $\,\lambda_x > \lambda_c\,$) or larger than one (for $\,\lambda_x < \lambda_c$). When $\,\lambda_x <0 \,$, we have $X_B > 1$ and sign$(X_A)$~=~sign$(\lambda_c)$. Lastly, in the case $\,\lambda_c <0 \,$, we find that $\,X_A < 0\,$ and $\,X_B \leq 1\,$ if and only if $\,\lambda_x \geq 0$.

Another possibility is the existence of a catastrophical ``point'' at infinity ($X \rightarrow - \infty\,$,$\;Y \rightarrow + \infty$), to which the system can collapse in some cases (the reciprocal point with positive $X$ and negative $Y$ can be shown not to exist). To illustrate such a case let's consider, for the sake of argument, $\,\lambda_x <0\,$, $\,\lambda_c >0\,$, $\,Y>0\,$ and \linebreak[4] $\,z = u = 0\,$. Then, the term $\,- 3\lambda_c \,Y\,$ is negative and we see from~\eqref{eq:autonomous} that $\,dX/dN\,$ will be negative for small enough positive $X$, and will stay so after $X$ becomes negative. At the same time, $\,dY/dN\,$ will be positive for large enough values of $\,\lambda_c$. A catastrophic decay thus ensues and $X \rightarrow -\infty\,$ in finite $N$.\footnote{In this case none of the energy densities are actually growing to infinity, but rather it is $H$ that is going to zero.}

Finally, as pointed out before, the system has three other points of interest. A baryonic point~$C$, a radiation dominated point~$D$, and a mixed matter dominated point~$E$, defined respectively as $\,(0,0,1,0)\,$, $\,(0,0,0,1)\,$ and $\,(0,Y,z,0)\,$. We find that point~$C$ is always a saddle point, point~$D$ may be either unstable or a saddle point depending on the values of $\lambda_x, \lambda_c$ and $w_x$, and that point~$E$ is a saddle point unless $|w_x|<\lambda_x$, in which case it is an attractor.

\subsection{Stability of the fixed points}

\begin{figure}
    \includegraphics[width=6.2cm]{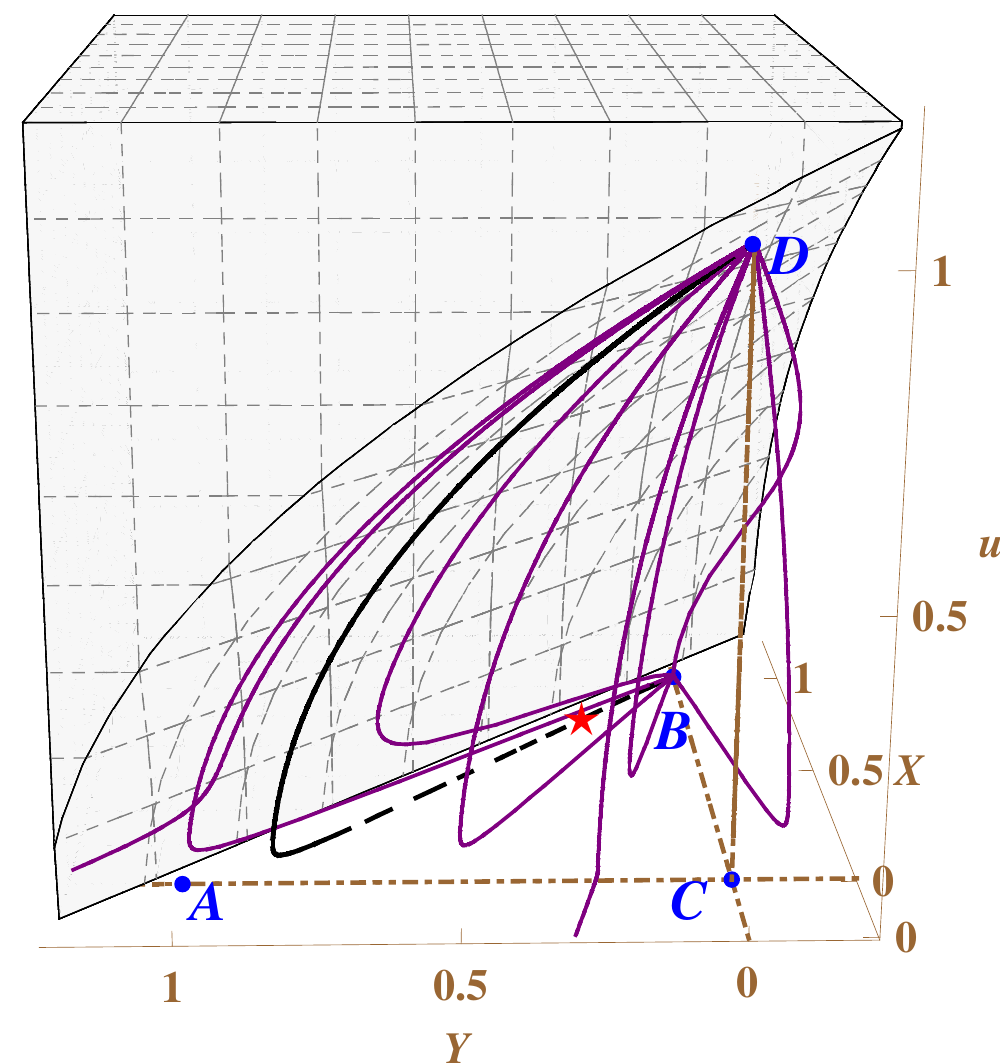}
    \includegraphics[width=6.2cm]{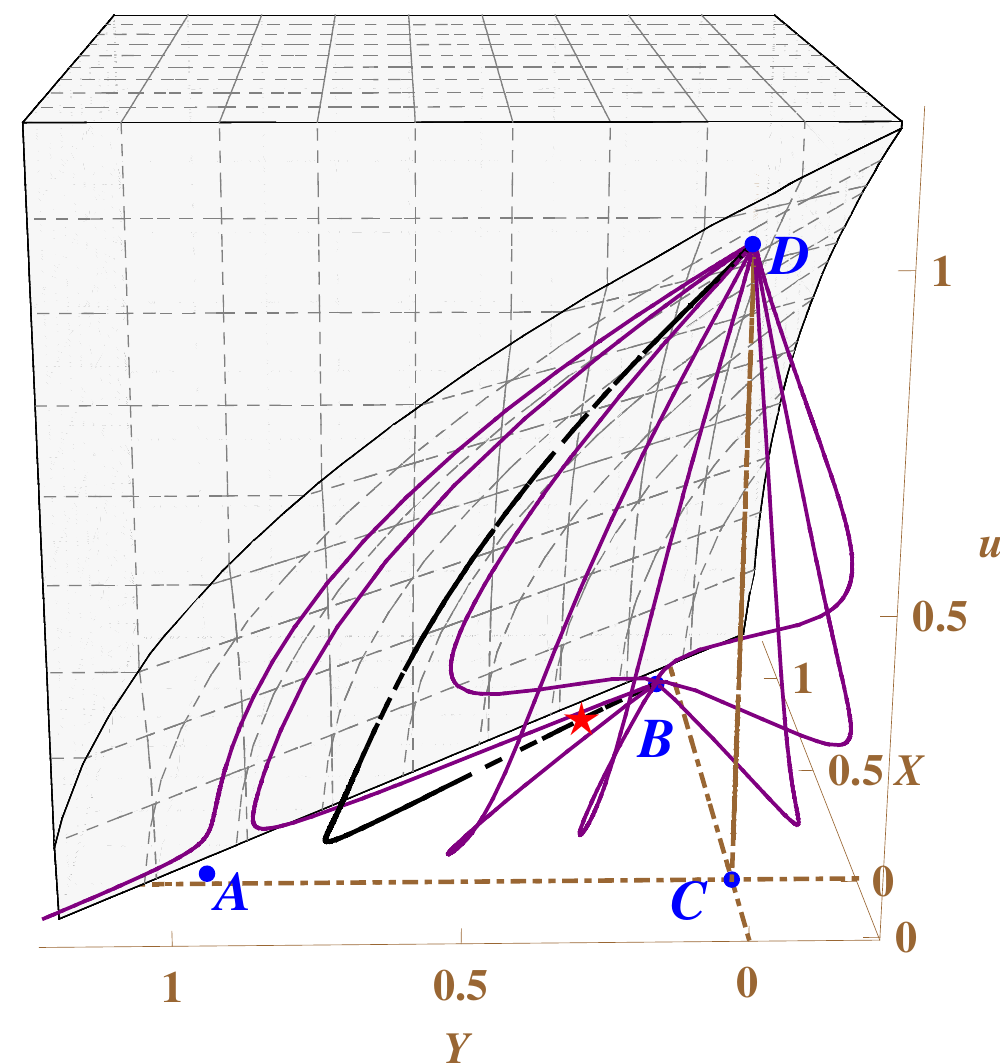}
    \includegraphics[width=6.2cm]{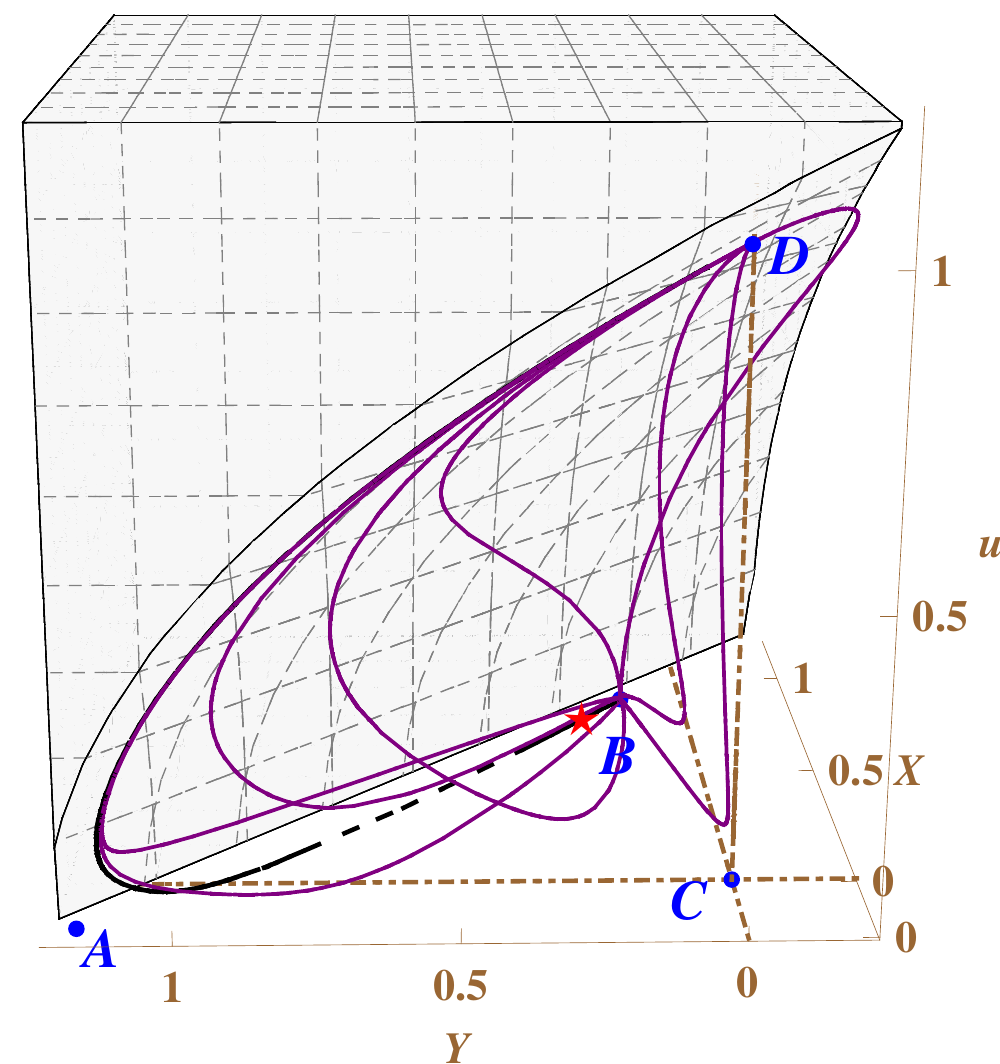}
    \caption{Phase space trajectories together with the fixed points $A$, $B$, $C$ and $D$. From top to bottom we have $\,\lambda_x=\lambda_c=0\,$, $\,\lambda_x=\lambda_c=0.04\,$ and $\,\lambda_x=-\lambda_c=-0.15$, while $w_x=-1$ for all three plots. Note that some trajectories cross either the $\,X=0\,$ or $\,Y=0\,$ surfaces. The gray volume represents the region where the constraint (\ref{eq:constraint-3d}) is violated. The black dashed line corresponds to a trajectory that passes through the present (observational) energy densities, which in turn is depicted by the red star.}
    \label{fig:phasespace}
\end{figure}

Table \ref{tab:fixed-pts} summarizes the conditions for existence and stability of all 6 fixed points. For the sake of clarity, we subdivide each point $A$ and $B$ into three, according to whether $X_{\rm fp}<0$, $0\leq X_{\rm fp}\leq 1$ or $X_{\rm fp}>1$.

Regardless of the positivity of $\lambda_x$ and $\lambda_c$, points $A_2$, $A_3$ and $B_1$ are always saddle points, whereas points $B_2$ and $B_3$ are always attractors. Point $A_1$ is a little trickier. It will be a saddle for either
\begin{equation} \label{eq:cond-saddle-XA1}
    \left\{
    \begin{aligned}
        &|w_x|(1+3\lambda_c) > \lambda_x - \lambda_c - \frac{1}{3}  &\mbox{ if }  \lambda_c \leq 0\,;\\
        &\frac{\left[\lambda_x - \lambda_c - \frac{1}{3}\right]}{(1+3\lambda_c)} < |w_x| \leq \frac{1}{9\lambda_c}
        &\mbox{ if } \lambda_c > 0\,.
    \end{aligned}
    \right.
\end{equation}
In all other cases, it will be unstable. In any case, whereas points $A_2$, $A_3$ and $B_1$ have $2$ negative (and $1$ positive) eigenvalues, point $A_1$ may have at most only one negative eigenvalue.

The baryonic point $C$ is a saddle point, but since we do not expect a baryonic dominated era, for reasonable initial conditions trajectories should not pass too close to it (although see section \ref{sec:toy-model}). As for the radiation dominated point $D$, even though it may technically be a saddle point (depending on the values of $\lambda_x$, $\lambda_c$ and $w_x$), in all physically relevant scenarios it will be unstable.

\subsection{Summary of fixed points} \label{summary}

Throughout this subsection we shall assume, for simplicity, that neither peculiar conditions~(\ref{eq:condition-a-b-w}) nor \linebreak[4] $\lambda_c = 0\,$ hold. We thus have a total of 4 fixed points plus a possible ``catastrophical abyss'' dubbed~$F$. Point~$D$ is the initial condition for our system. Proximity to point~$C$ is better avoided in order to cope with observations. Points~$A$ and~$B$ are the main focus of interest: the former, a candidate for a saddle MDE and the latter, a candidate for the present accelerated epoch. Point~$A$ and~$B$ will usually be a saddle and an attractor, respectively, but there are exceptions.

In this model, contrary to the particular case~\cite{Olivares07}, we have {\large $\frac{X_A}{Y_A} \neq \frac{Y_B}{X_B}$}. In fact, we have $X_A-Y_B = ${\large $\frac{\lambda_x -\lambda_c}{w_x}$} instead of zero. This is a desirable result, for it allows the accelerated fixed point to have a dark energy density closer to the present value. Put in another way, when one adjusts the MDE fixed point found in~\cite{Olivares07} to cope with observations, the accelerated point acquires a ratio $R=\rho_x / \rho_c \gtrsim 45$ (see sections~\ref{sec:obs-constraints} and~\ref{sec:toy-model}), while present observations state that $R_0 \approx 3.4$. We will come back to this issue on section~\ref{sec:toy-model}, where we investigate possible improvements when the two coupling constants differ.

A sample of possible trajectories are depicted in figure \ref{fig:phasespace} for three different cases, all of which have $w_x$ set to $-1$: $\lambda_x = \lambda_c = 0$ ($\Lambda$CDM), $\lambda_x = \lambda_c = 0.04\,$ and $\,\lambda_x = -\lambda_c = 0.15$. The gray volume denotes the forbidden region, i.e., the one for which the constraint~\eqref{eq:constraint-3d} is violated. The black dashed curve represents a trajectory that passes through the present (observational) energy densities, which in turn is depicted by a red star. Note that, as pointed out before, in all cases it is possible for trajectories to cross either the $\,X=0\,$ or $\,Y=0\,$ surfaces. Had we chosen $\frac{1}{H}\sqrt{\frac{\rho _{x/c}}{3}}$ as variables, we would see trajectories hitting such surfaces but artificially not being able to cross them. This is specially important for models with negative values of $\lambda_x$ and/or $\lambda_c$, for which the fixed points $A$ and $B$ may have either $\,X_{\rm fp}<0\,$ or $\,Y_{\rm fp}<0$.

\section{Possibility of two scaling epochs} \label{sec:2epochs}

We now analyse the feasibility of a duo-scaling cosmology, in which the MDE is assured by the existence of a suitable saddle point and the present acceleration by the existence of an appropriate final attractor. From table~\ref{tab:fixed-pts} it is clear that we need point $A$ to be such a saddle, with $X_A \lesssim 0.1\,$, and point $B$ to be the attractor, with $X_B \gtrsim 0.7\,$. In other words, points $A_3$ and $B_1$ are excluded. Point $B_3$, in turn, is not very interesting for it would mean that: (i)~the universe is now in a transient phase between the two scaling epochs, which inevitably requires a more precise tuning of the parameters; and (ii)~the dark matter energy density would become negative in the future. The former is a crucial feature, as it automatically prevents solution to the CP in the way we defined it. We shall thus consider only the $B_2$ flavour of point $B$ and consequently require $\,\lambda_x \geq 0$. In a nutshell, we limited ourselves to only two possibilities:
\begin{itemize}
    \item Point $A_1$ followed by $B_2$ (requiring $\lambda_c <0$);
    \item Point $A_2$ followed by $B_2$ (requiring $\lambda_c >0$).
\end{itemize}

To make point $A_1$ a viable saddle we need $\lambda_c$ to be negative, since conditions~(\ref{eq:condition1-on-w}) and~(\ref{eq:condition2-on-w}) are mutually exclusive. Also, from eq.~(\ref{eq:cond-saddle-XA1}) and the fact that $\,\lambda_x \geq 0\,$, we find that $\lambda_c$ must be larger than $-1/3$. But even $\,-1/3 \leq \lambda_c < 0\,$ might not be enough, since point $A_1$ has at most one negative eigenvalue and, as such, it is unclear if it would have a large basin of attraction as required to address either the fine-tuning of initial conditions (SIC) or the CP. In other words, the corresponding MDE might be short and it may lead to smaller values of $\Delta$.

Points $A_2$ and $B_2$ are easier to reconcile. In fact, a sufficient condition for the system to have the two scaling regimes we need in other to attack the coincidence and initial condition problems is $\lambda_x , \lambda_c >0$ and $w_x$ satisfying the constraint~(\ref{eq:condition1-on-w}). However, as will be shown in the next section, observational constraints put limits on the coupling strength which are tighter in this case than in the one with $\,\lambda_c < 0.$

Following these guidelines, we develop a quantitative analysis in section~\ref{sec:fine-tuning} for some of the observationally allowed values of the parameter space. We also investigate the possible improvements on the range of initial conditions that eventually evolves to today's observable universe (i.e., the DEICP).

\section{Observational Constraints} \label{sec:obs-constraints}

In order to probe quantitatively the permissible range of values of the two coupling constants, we submit the model to two different observational tests: type~Ia supernovae and the so-called CMB shift parameter. The former relies on the well accepted hypothesis that this kind of astronomical object is a standardizable candle, and consists of comparing their distance moduli at different redshifts to the ones calculated from the model. The latter can be considered a good first approximation of the full CMB analysis (specially when used in conjunction with the acoustic peak scale~\cite{Elgaroy07}), which would require setting up the first order perturbation equations for the model and employing one of the established CMB codes, such as CAMB or CMBEASY.

We made use of the combined Essence, Hubble, SNLS and nearby supernovae catalog as compiled by~\cite{Davis07}\footnote{This catalog has the well-known feature of agreeing with $\Lambda$CDM within $1\sigma$. Here we are not advocating its use over other SNIa sets, but the reader should bear this in mind when analysing our confidence levels contours; to wit, the inclusion of interactions do not change this picture.}, for a total of 192 supernovae, and of the WMAP 3-year result for the shift parameter~\cite{Wang07} (to wit, $1.70\,\pm\,0.03$). We employed a grid-based method to compute the $\chi ^2$ for different values of $\,\lambda_x, \,\lambda_c, \,w_x\,$ and $\,\Omega_{c0}$. The present values of the baryonic energy density $\Omega_{b0}$ and the radiation energy density $\Omega_{r0}$  were held fixed respectively at $0.042$ and $4.2 \times 10^{-5}$, which are the best fit values of the combined WMAP3 and SDSS observations~\cite{Spergel07}. The value of $h$ was marginalized analytically. For $\Omega_{c0}$ we assumed a gaussian prior equivalent to WMAP3+SDSS observations, i.e., with a mean at 0.22 and a standard deviation of 0.03. We also considered a top-hat prior in the range $0.14 < \Omega_{c0} < 0.30$, but results were very similar.

\begin{figure}[t]
    \includegraphics[width=8.1cm]{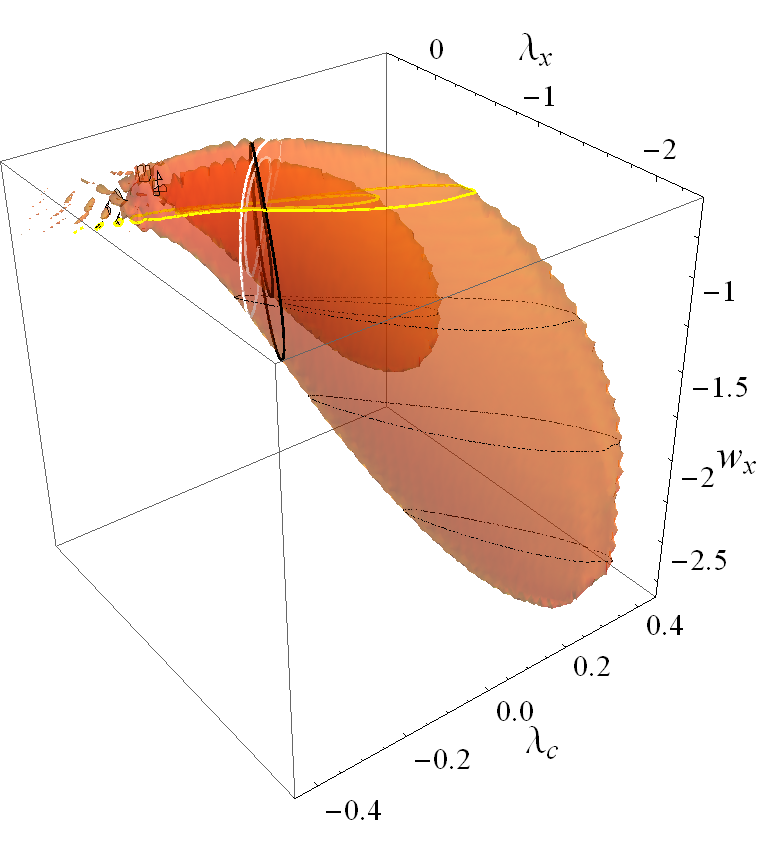}
    \caption{Combined results from supernovae and shift parameter tests. The volumes represent $1$ and $2\sigma$ confidence levels, marginalized over $\Omega_{c0}$ with a gaussian prior based on WMAP3+SDSS results. The vertical contours are drawn at $\lambda_x = 0$ (white) and $\lambda_c=0$ (black). The yellow horizontal cut is made at $w_x = -1$ (see figure \ref{fig:Contours-wx=-1}), and further dashed cuts are made to aid the eye at $w_x = -1.5$, $w_x = -2.0$ and $w_x = -2.5$.}
    \label{fig:Combined-3D}
\end{figure}

\begin{figure}[t]
    \includegraphics[width=8cm]{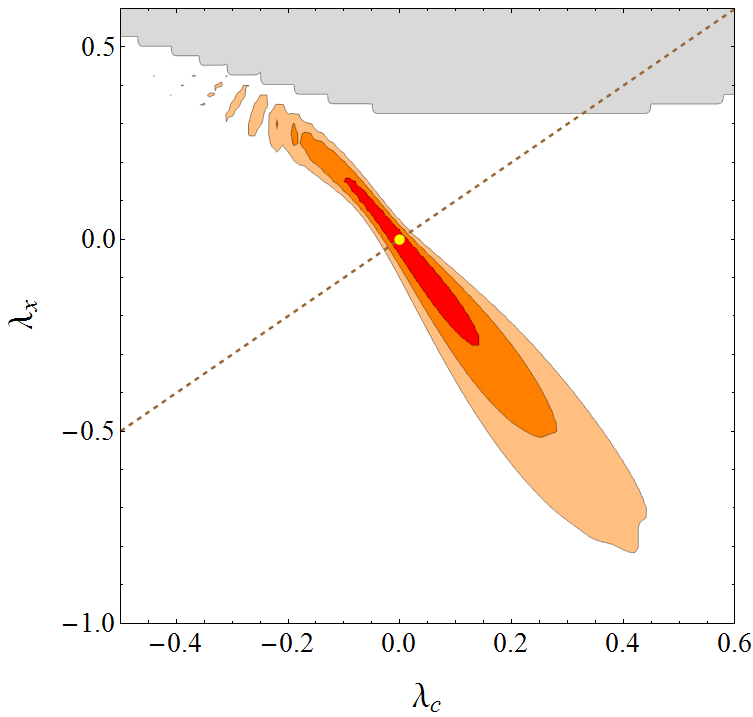}
    \caption{$1$, $2$ and $3\sigma$ confidence level contours for $w_x$ fixed at $-1$. The brown dashed line shows the particular case of $\,\lambda_x~=~\lambda_c\,$, as considered in~\cite{Chimento03}, and the yellow dot stands for the $\Lambda$CDM case. The gray area on top is the region for which the system enters the ``catastrophical abyss'' characterized by the fixed point $F$, which in practice means that $\,H(z^\ast) = 0\,$ for some $z^\ast < \,z_{\rm recombination}$.}
    \label{fig:Contours-wx=-1}
\end{figure}

\begin{figure*}[t]
    \includegraphics[width=14.5cm]{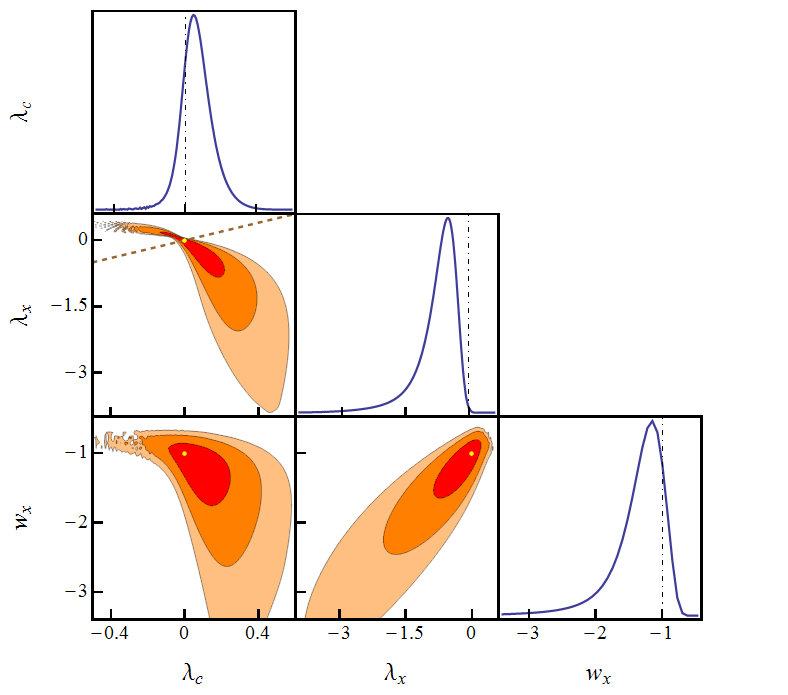}
    \caption{Marginalized $1$, $2$ and $3\sigma$ contours for each of the three different model parameters, and the one dimensional likelihoods. All priors were taken to be flat. The black vertical lines on the one dimensional plots and the yellow dots on the contour plots represent $w_x = -1$, $\lambda_c =0 $ and $\lambda_x =0$ respectively. The brown dashed line on the bottom middle plot stands for the $\,\lambda_x = \lambda_c\,$ case.}
    \label{fig:Marginalized-Plots}
\end{figure*}

\begin{figure}[t]
    \includegraphics[width=8.3cm]{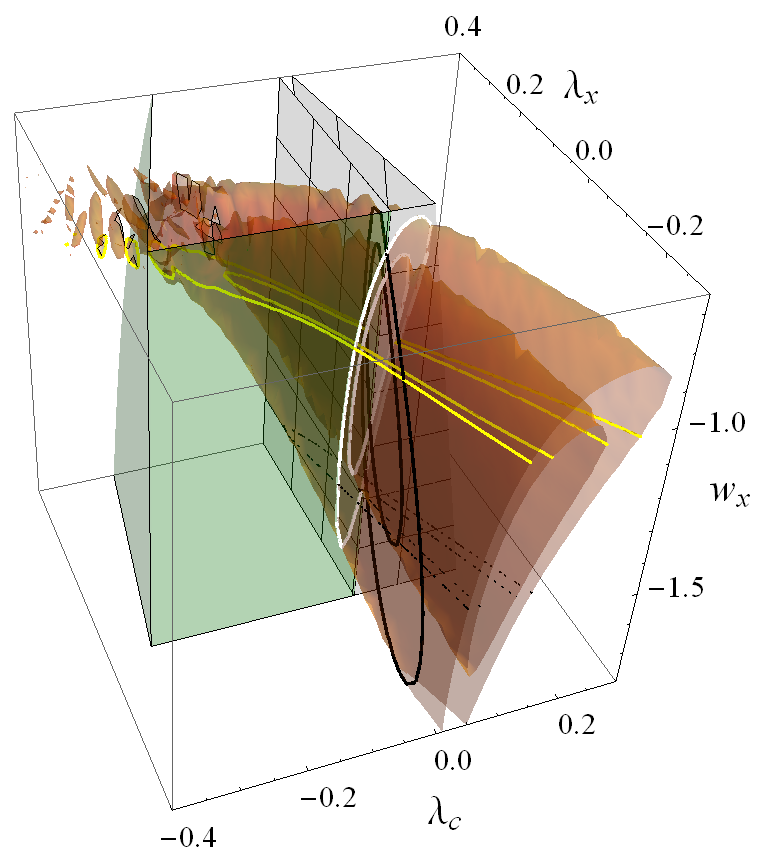}
    \caption{Intersection of the $1$ and $2\sigma$ contours with the regions of the parameter space that allow for a duo-scaling regime. Inside the green transparent region we have a $A_1-B_2$ type of scaling, whereas inside the gray opaque checkered region the scaling is of the type $A_2-B_2$. The gray borders are drawn in such a way to guarantee that $X_{A_2} < 0.1$, so as not to compromise the formation of structures during the MDE. The 2D contours are the same as in figure \ref{fig:Combined-3D}.}
    \label{fig:Duo-Scaling}
\end{figure}

Figure \ref{fig:Combined-3D} depicts the $1$ and $2\sigma$ confidence levels for the gaussian prior and contains much information about the model. It shows that, in absolute terms, negative values of $\lambda_x$ can be many times larger than positive ones. It is also clear that neither test disfavours phantom values of $w_x$, and that the allowed region in the $\lambda_x > 0, \,\lambda_c > 0$ quadrant is quite small. A two dimensional ``cut'' at $w_x = -1$ is shown in figure~\ref{fig:Contours-wx=-1} and gives a good idea of the range of values $\lambda_x$ and $\lambda_c$ can take as well as the overall alignment of the contours with the line $\lambda_x = - k \lambda_c$ where $k \sim 2$. Technically it is not a cut from figure~\ref{fig:Combined-3D} but rather the contours for a two parameter model with $w_x$ set at $-1$ from the beginning. In practice the only difference is that the contours in figure~\ref{fig:Contours-wx=-1} are smaller than the yellow ones in figure~\ref{fig:Combined-3D}. Figure~\ref{fig:Marginalized-Plots} shows different two dimensional contours derived from the higher dimensional likelihood, assuming flat priors on all 3 model parameters, as well as the one dimensional likelihood for each one. The use of flat priors allows the interpretation of all these six plots as projections of the higher dimensional ones, thus helping visualize figure~\ref{fig:Combined-3D}. Finally, figure \ref{fig:Duo-Scaling} merges the three dimensional plot with the regions of the parameter space that exhibit the duo-scaling regime. In the interior of the green transparent region we have a $A_1-B_2\,$ duo-scaling, whereas in the interior of the gray opaque checkered volume we have a $A_2-B_2\,$ one. In addition to the existence and stability of the fixed points $A_2$ and $B_2$, the gray borders are drawn in such a way to guarantee that $X_{A_2} < 0.1$. The idea is that dark energy needs to be sub-dominant during the MDE in order not to interfere too much with the formation of structures. In any case, as can be seen from the plot, a fixed point $X_{A_2}>0.1$ is excluded at over~$2\sigma$.

The major focus of previous works which adopted similar coupling models has been in the positive $\lambda_x,\,\lambda_c$ quadrant. It is now clear that such a region is a severely limited piece of the allowed parameter space. From figure~\ref{fig:Contours-wx=-1} we see that if we limit ourselves to, say, $\,\lambda_{x}=\lambda_{c}>0\,$ as done in~\cite{Olivares06}, we must have $\lambda_{c} < 0.014$ $(0.020)$ at \linebreak[4] $2\sigma$ ($3\sigma$) confidence levels. If we consider negative $\lambda_{c}$, on the other hand, the coupling constants can be as large as 0.30 $(0.42)$ and $-0.20$ $(-0.35)$ at $2\sigma$ ($3\sigma$) for $\lambda_{x}$ and $\lambda_{c}$, respectively. That is over an order of magnitude increase on absolute values. We will exploit such freedom in what follows and investigate the feasibility of models with negative values of $\lambda_{c}$.

\section{Fine-tuning and the Initial Conditions Problems} \label{sec:fine-tuning}

\begin{figure}[t]
    \includegraphics[width=7.8cm]{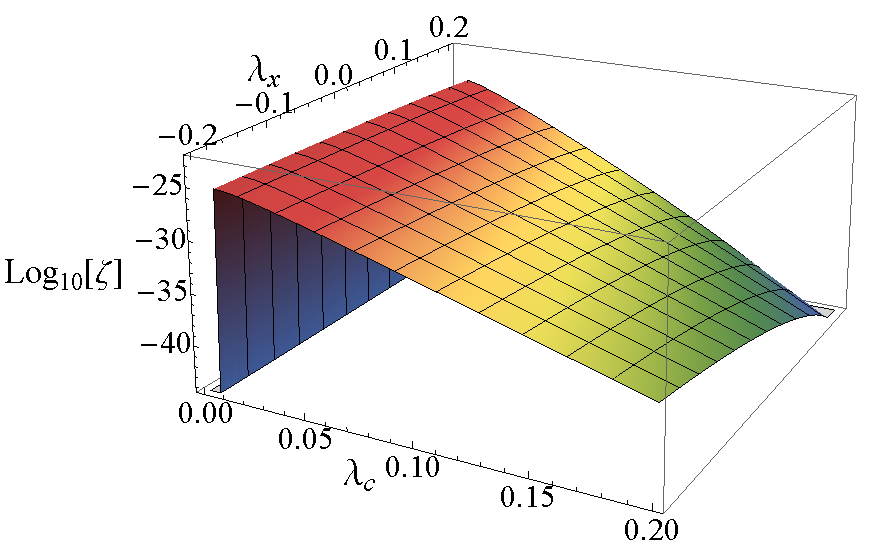}
    \includegraphics[width=7.5cm]{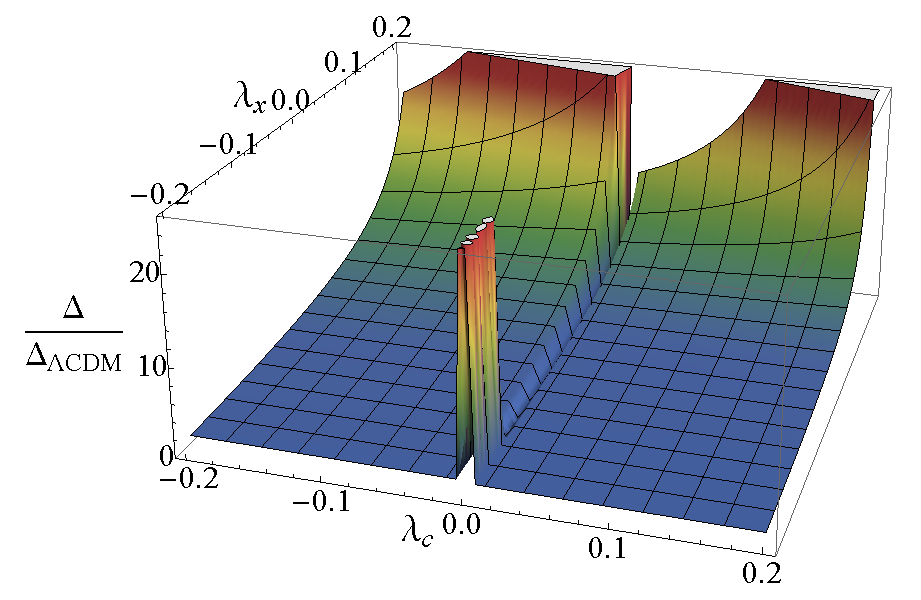}
    \caption{The DEICP variable $\zeta$ (top) and SIC variable $\Delta$ (bottom), plotted as a function of the coupling constants $\lambda_x$ and $\lambda_c$. Here we set $\,w_x=-1\,$. The improvements in $\zeta$ range between 55 and 75 orders of magnitude ($\zeta_{\Lambda \textrm{CDM}} \sim 10^{-100}$), while $\Delta$ is between 1 and 30 times the $\Lambda \textrm{CDM}$ one. The abrupt decline of $\zeta$ close to $\,\lambda_c=0\,$ is expected (see section~\ref{sec:toy-model}). The width of the valley at $\,\lambda_c=0\,$ in the plot of $\Delta$ is grossly exaggerated here for clarity purposes. In fact, the drop only starts at $\left|\lambda_c\right| \lesssim 10^{-60}$. For $\lambda_c < 0$, $\zeta$ is negative, and thus looses its relevance since the DEICP is no longer applicable.}
    \label{fig:Initial-Conditions}
\end{figure}

As pointed out before, there are two ways in which any model that aims to address the DEICP with small SIC may offer improvements. One is to increase the value of the its initial DE energy density at some given time in the past, or more precisely, to increase the ratio $\zeta$ of energy densities between dark energy and radiation at the end on inflation. The other is to increase the range~$\Delta$ which will evolve to the observed values of $\,\Omega_{x0}$. The former may help conciliate the dark energy component, under a field interpretation, with our present understanding of the Standard Model of particle physics or of its extensions. The latter is a direct measure of the basin of attraction of the MDE saddle point and, therefore, of the reduction of fine-tuning required.

A model which undergoes a duo-scaling regime is expected to be less sensitive to initial conditions due to the presence of the MDE saddle point $A$. The amount of loosening is nonetheless unclear from the previous arguments. One needs therefore to assess the range of allowed initial conditions for, say, the energy density of dark energy for different values of the parameter space allowed by observations and compare it with the range of initial conditions in the $\Lambda$CDM model. This is done in figure~\ref{fig:Initial-Conditions}, where for $\,w_x=-1\,$ we plot the values of $\,\Delta / \Delta_{\Lambda \textrm{CDM}}\,$ and $\zeta$ by computing the range of initial values of $\rho_x$ that evolve to present day observations ($X_0 = 0.74 \pm 0.03$). As noted in section~\ref{sec:intro}, we take our initial values at the end of inflation, more precisely at $z=10^{26}$. For reference, $\,\Delta_{\Lambda \textrm{CDM}} \simeq 0.08\,$ (i.e., initial conditions for $\rho_x$ may vary $\sim 8\%$, which is of course what we expect since in $\Lambda \textrm{CDM}\,$ $\rho_x$ is constant and that is its current uncertainty) and $\,\zeta_{\Lambda \textrm{CDM}} \simeq 9 \times 10^{-101}$. We see that when interactions are at work, improvements of a factor between 1 and 30 for~$\Delta$ and of dozens of orders of magnitude for $\zeta$ (as long as $\lambda_c \neq 0$) are possible. For small but positive values of $\lambda_c$ (starting around $10^{-4}$), $\zeta$~drops abruptly to the $\Lambda$CDM value. We remark that the negative $\lambda_c$ region has negative $\rho_{xi}$, and thus in this section $\zeta$~looses its importance since the DEICP is no longer applicable (remember that we are identifying the DEICP with equipartition arguments).  Note that the valley close to $\,\lambda_c = 0\,$, and the thin peak at its base ($\lambda_x \simeq - 0.18$) in the plot of $\Delta$ were artificially widened in figure~\ref{fig:Initial-Conditions} for clarity purposes. Both are actually much thinner than depicted, and only exist for $\left| \lambda_c \right| \lesssim 10^{-60}$. As a matter of fact, this peak at the bottom is spurious: $\rho_{xi}^{\rm max}$, the denominator of $\Delta$, is getting very close to zero there. As predicted in section~\ref{sec:2epochs}, the region of positive $\lambda_x > 0$ is less sensitive to initial conditions, which is due to the presence of the duo-scaling regime.

Two important remarks are in order here. First, as far as the DEICP is concerned, it is much more sensitive to $\lambda_c$ than $\lambda_x$. In fact, the highest possible values of $\zeta$ are obtained for very small and positive $\lambda_c$. Stronger couplings with DM only make the DEICP worse, where as no values of $\lambda_x$ allowed by the observational tests applied make a difference when $\lambda_c \ll 1$. Second, in the region of positive $\lambda_x$, which is the interesting one since in allows the existence of both scaling regimes, the SIC is considerably smaller than in $\Lambda$CDM, which is an advantage for any phenomenological approach. In particular, the region around $\,\lambda_x = 0.2\,$ and $\,\lambda_x = -0.1$, which show improvements in $\Delta$ by factors of $10-30$, turn out to be the most promising one to solve the CP, as will be discussed in the next section.

\section{A two-variable duo-scaling toy model} \label{sec:toy-model}

Inspired by the results on the preceding sections, we look for a toy model which at the same time alleviates the DEICP, do not has a large SIC, undergoes a duo-scaling regime and satisfies the observational tests carried out in this work. A good starting point is to select a model which has an analytic solution. As mentioned in section~\ref{sec:DIM}, the best candidate is the one in which \begin{equation} \label{eq:toy-model}
    \,w_x=-1-\lambda_x \,.
\end{equation}
In this case, we can write the ratio $R$ between $\rho_x$ and $\rho_c$ as
\begin{equation} \label{eq:toy-ratio}
    R \,=\, \frac{(S+L) R_0 - 2 \lambda_c  + (1+z)^{3 S} \left[ (S -L)R_0 + 2\lambda_c \right] }{S - L + 2 R_0 \,\lambda_x + (1+z)^{3 S} \left[(S + L) - 2 R_0 \,\lambda_x\right]},
\end{equation}
where $\,S \equiv \sqrt{1 + \lambda_c(-2 - 4\lambda_x + \lambda_c)}\,$, $\,L\equiv 1 - \lambda_c\,$ and $\,R_0\equiv \rho_{x0}/\rho_{c0} \simeq 3.4\,$. To better understand the influence of the two coupling constants it is fruitful to write down the above expression in limits $\,z \rightarrow \infty\,$ and $\,z \rightarrow -1$. Since we will only be considering coupling constants which are reasonably smaller than unity, we will neglect all but the two lowest order terms in $\lambda_x$ and $\lambda_c$ in \eqref{eq:toy-ratio}. We thus get
\begin{align} \label{eq:limits-of-R-inf}
    \lim_{z \rightarrow \infty} R &\,=\, \lambda_c \left( 1 + \lambda_c - R_0\,\lambda_x \right) + O(\lambda^3)\,, \\
    \lim_{z \rightarrow -1} R &\,=\, \frac{R_0 + \lambda_x\, - \lambda_c \left(1 + R_0\right)}{\lambda_x} + O(\lambda)\,. \label{eq:limits-of-R-1}
\end{align}
We see that for $\,\lambda_c = 0$, $R$ goes asymptotically to zero at high redshifts. If $\lambda_c$ in nonzero we get (unless very specific values of the couplings are chosen to cancel the term proportional to $1+z$ in the numerator) a constant, nonzero value for $R$ as $z$ goes to infinity. On the other extremity, as $\,z\rightarrow-1$, we see that setting $\,\lambda_x = 0\,$ means that the ratio of energy densities will grow indefinitely in the future. This gives a physical intuition on the roles of both $\lambda_x$ and $\lambda_c$ when they assume nonzero values. The former is needed to ensure a late-time scaling, i.e., to keep $R$ small, while the latter ensures a nonzero value of $R$ at early times. Put in another way, large values of $\lambda_x$ help solve the CP, while large values of $\lambda_c$ alleviate the DEICP. Referring back to section~\ref{sec:fpoints}, this could also be seen from the fact that $\,X_{A_2} = 0\,$ when $\,\lambda_c = 0\,$ and that $\,X_{B_2} = 1\,$ when $\,\lambda_x = 0$.

We now need to select particular values for $\lambda_x$ and $\lambda_c$ (remember that we are assuming $\,w_x = -1 -\lambda_x$) that passes the proposed observational tests and gives a final value $R_{-1}$ of $R$ which is not far from the present one. From figures~\ref{fig:Combined-3D} and~\ref{fig:Contours-wx=-1} we see that the lowest asymptotic value of $R$ allowed in the positive coupling constants quadrant is $R_{-1} \simeq 32\,$ ($R_{-1} \simeq 45\,$ if $\,\lambda_x=\lambda_c$), while for negative $\lambda_c$ we can get as low as $\,R_{-1} \simeq 5$. \linebreak[4] In any case, the coincidence problem is at best only alleviated, not solved, as we would still be somewhat far from these values. Based on these considerations, we select $\left\{\lambda_x = 0.18, \,\lambda_c = -0.08,\,w_x = -1.18\right\}$ and $\left\{\lambda_x = \lambda_c = 0.014,\,w_x = -1.014\right\}$ as examples of our toy model. In the former, early DE has negative energy density, and thus it is not a candidate for solving the DEICP. The latter has positive dark energy throughout the whole history, but exhibits a worse CP.

\begin{figure}[t]
\includegraphics[width=8cm]{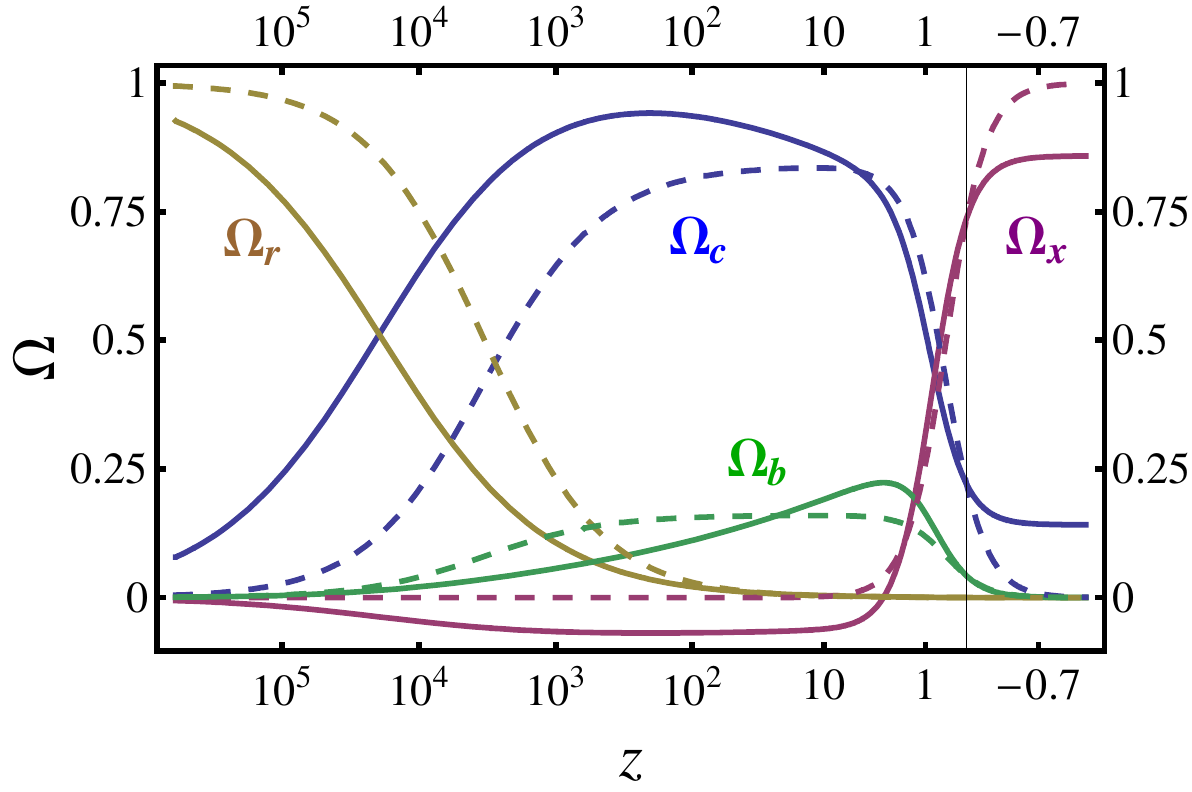}
    \caption{Evolution of all four energy densities as a function of redshift. From left to right, the dominant component is, in order: radiation, dark matter and dark energy. Baryons are always sub-dominant. The dashed lines correspond to the $\Lambda$CDM model and the solid lines to the case $\lambda_x~=~0.18$, $\lambda_c~=~-0.08$ and $w_x~=~-1.18$. Note that matter-radiation equality is pushed back to $\,z \simeq 20000$, and that DE has negative energy density for $z > 2$.}
    \label{fig:Toy-Model}
\end{figure}

\begin{figure}[t]
\includegraphics[width=8cm]{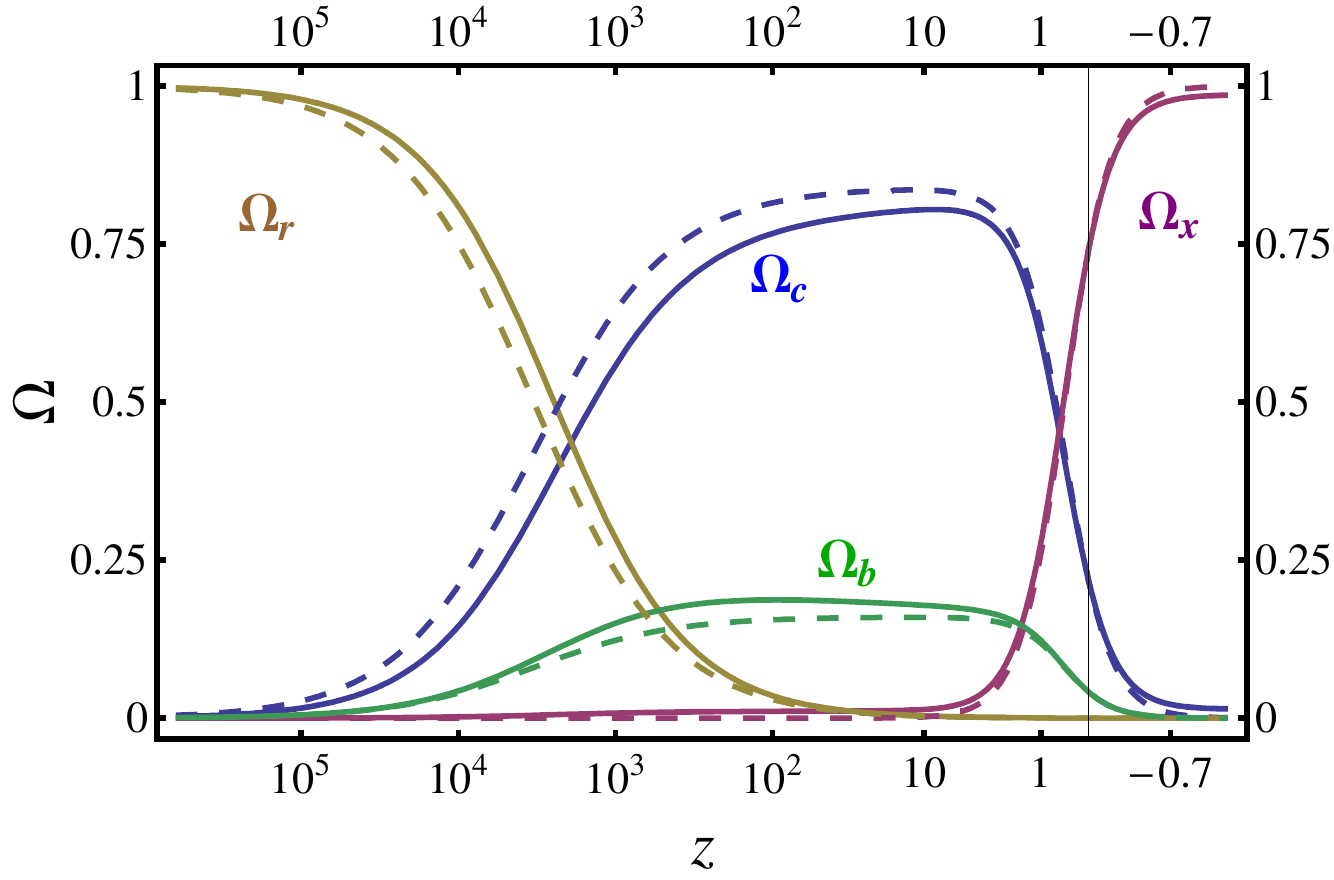}
    \caption{Same as figure \ref{fig:Toy-Model} for the case where $\lambda_x\,=\,\lambda_c\,=\,0.014$, $w_x\,=\,-1.014$. Here the evolution resembles more closely that of $\Lambda$CDM, and in particular the DE energy density is always positive.}
    \label{fig:Toy-Model-2}
\end{figure}

In figure~\ref{fig:Toy-Model} we plot the different energy densities as a function of redshift for the first chosen set of parameters and compare them with $\Lambda$CDM. The present values of the different $\Omega$ are set by the WMAP3+SDSS best fit. Note that $\rho_x$ is negative throughout the entire matter dominated era, which helps to increase its duration. Another important feature of this example is that the radiation-matter equipartition occurs much earlier, at $z \simeq 20000$, and therefore it is possible that different observational tests such as a full CMB analysis would rule out such an anticipation of the MDE. In fact, fitting the observed matter power spectrum would be a challenge since its peak is roughly estimated by $k_{\rm eq}$, the wave number of perturbations which enter the horizon at matter-radiation equality, and that would be shifted to smaller scales by a factor of $\simeq 6.1$ (in units of $h$ $\times$ distance$^{-1}$). This example has small SIC, as can be seen from figure~\ref{fig:Initial-Conditions}. Figure~\ref{fig:Toy-Model-2} is the same for the second example, for which $\,\lambda_x = \lambda_c= 0.014$. In this case, the DEICP does apply and we get $\zeta \simeq 3 \times 10^{-26}$ which should be compared to $\zeta_{\Lambda \textrm{CDM}} \simeq 9 \times 10^{-101}$. Note that in this case we are farther from the final accelerated attractor than in the previous one. In fact, $R_{-1} = 6.1$ in the first example and $R_{-1} = 70.4$ in the second one.

\begin{figure}[t]
\includegraphics[width=8cm]{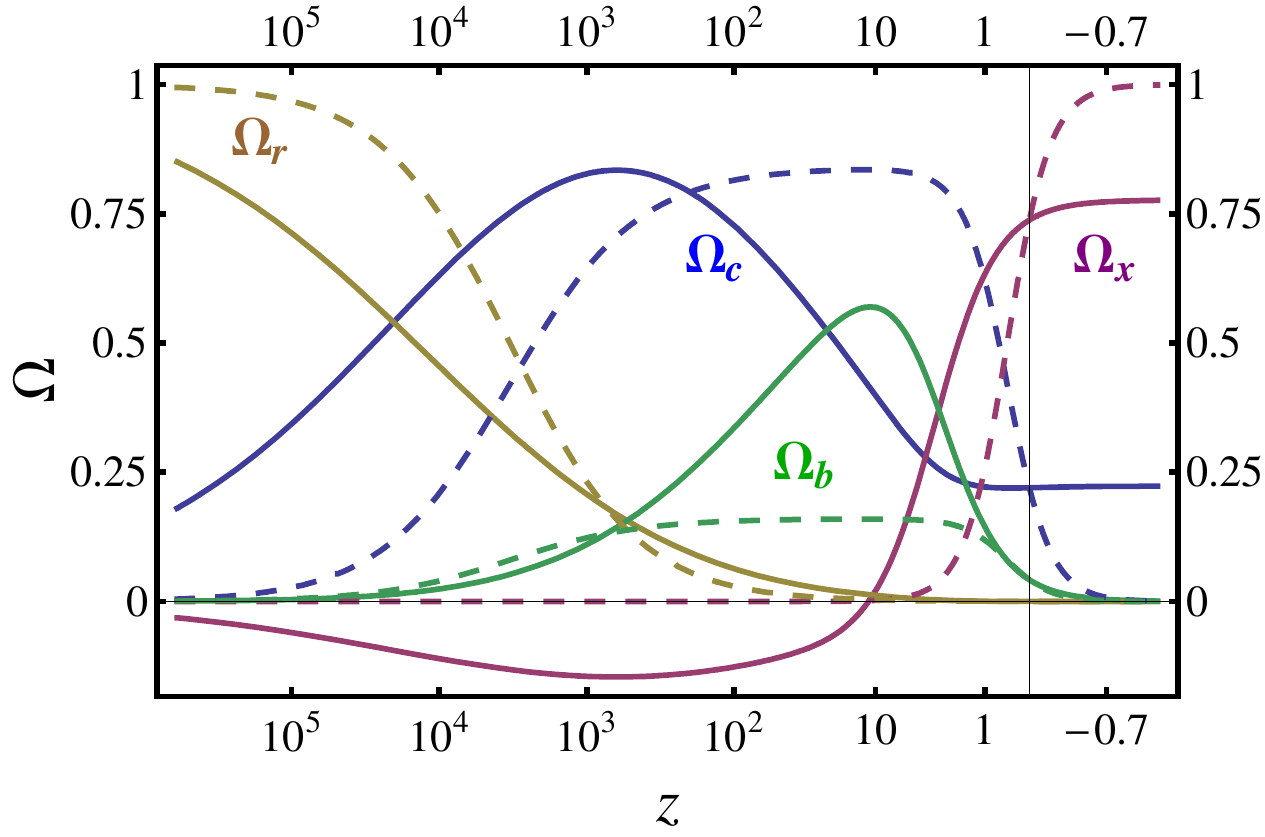}
    \caption{Same as figure \ref{fig:Toy-Model} for the case where $\lambda_x=0.23$, $\lambda_c=-0.14$, $w_x=-0.85$. Again, matter-radiation equality is pushed back to $\,z \simeq 20000$. Note that for a not-so-short redshift interval, baryons become the (marginally) dominant component. In this scenario, we have already reached the accelerated attractor ($R_{-1}$ within $4\%$ of $R_0$) and thus, by our definition, solved the CP.}
    \label{fig:Toy-Model-3}
\end{figure}

We conclude this section by noting that it seems to be possible to get $R_{-1}$ closer to $R_0$ for regions of the parameter space that do not allow analytic solutions. Inspired by eq.~\eqref{eq:limits-of-R-1}, we seek large values of $\lambda_x$, and from figure~\ref{fig:Combined-3D} we see that larger $\lambda_x$ are allowed when $\,w_x > -1$. Probing that region, we found a good candidate for $\left\{\lambda_x = 0.23 ,\,\lambda_c = -0.14,\,w_x = -0.85\right\}$. Figure~\ref{fig:Toy-Model-3} depicts the evolution of the model for these parameter values, which give $R_{-1} = 3.48$. This is within $4\%$ of the present (WMAP3+SDSS) value, and one could argue that this solves the CP. Nevertheless, once again the early equality might jeopardize the model (here $k_{\rm eq}$ is shifted by a factor $\simeq 6.6$). Another aspect we must point out is that this region of the parameter space is a rather sensitive one. This is reflected in the wiggles and discontinuities on the tip of the contours of figure~\ref{fig:Combined-3D}. What this means is that small changes in the parameter values are ``amplified'', i.e., may result in substantially different evolutions. For instance, for $\,\lambda_x = 0.235\,$ dark matter has negative energy density for $z>8$; for $\,\lambda_x = 0.20\,$ the matter-radiation equality is pushed back to $z>10^5$. This can be problematic if one tries to extend our analysis for varying $w_x$ and/or varying $\lambda_x$, $\lambda_c$ scenarios.

\section{Conclusions}

In this paper we tried to clarify some of the different aspects of cosmological fine-tuning and their relation to dark interactions. With this goal in mind, we proposed the use of three variables to quantify each of most common issues of cosmological models: the CP, the DEICP and the SIC.
By restricting our parameters through the use of both supernovae and the CMB shift parameter and through the requirement of a duo-scaling cosmology, we greatly limited our parameter space.

Applying the proposed variables to our Dark Interactions Model, which has some distinct features such as negative DE energy densities, we found that each coupling constant is related to a different regime. The CP can only be solved for large values of $\lambda_x$, while larger values of $\lambda_c$ guarantee higher amounts of DE in the early universe, and thus relate to the DEICP. We also found that nonzero values of $\lambda_c$ also give rise to smaller SIC for any value of $\lambda_x$. We thus investigated two examples of a toy model class (characterized by eq.~\eqref{eq:toy-model}) for which there exists analytic solutions so as to gain some intuition on how to address these different problems. For both of them the CP and DEICP were only at best alleviated, since the lower value of $R_{-1}$ obtained was $6.1$ and the higher value of $\zeta$ was $\sim10^{-26}$. Even if not a proper solution, the latter represent a $74$ order-of-magnitude improvement over the $\Lambda$CDM scenario. It is possible that a solution to the DEICP requires a third scaling solution: a scaling between dark energy and radiation. Such regime would only be achieved if we shed light on the dark interactions, including radiation in our coupling scheme.

It is also noteworthy the fact that, based on the examples shown, solutions to the CP (in our class of models) seem to require DE with negative energy density in the past. This may be a hint that modified gravity theories, which can more easily accommodate negative (effective) energy densities, might have an upper hand at explaining this issue. In fact negative-energy fields spreading over large spatial regions  might yield a number of exotic phenomena (see~\cite{Fewster98} and references therein).

Finally, we concluded that $R_{-1} \simeq R_0$ only occurs in a rather turbulent region of the parameter space, for which $w_x$ is not very negative and $\lambda_x$ is large enough. Since in this case the dark matter-radiation equality differs substantially from that of $\Lambda$CDM, it remains to be seen if further observational tests (specially those involving perturbations) will rule it out. In particular, coping with the shift in the peak of the matter power spectrum will be a challenge for this model. Should the perturbation equations turn out to be considerably different than those of the concordance model and the matter power spectrum fitted, then as sensitive as the model might be to its parameters in this region, compared to $\Lambda$CDM or quintessence no real fine-tuning will be necessary in order to solve the coincidence problem.

\section*{ACKNOWLEDGEMENTS}

M.Q., S.E.J., R.R.R.R. and I.W. are supported by the Brazilian research agency CNPq. M.O.C. is supported by the Rio de Janeiro research agency FAPERJ. \linebreak The authors would like to thank L. Raul Abramo for fruitful discussions.

\end{document}